\documentclass[superscriptaddress,showpacs,twocolumn,floatfix]{revtex4-1}

\usepackage{amsmath,amsthm}
\usepackage{IEEEtrantools}
\usepackage{algpseudocode}
\usepackage{algorithm}
\usepackage{graphicx}

\pdfoutput=1 

\usepackage{soul}
\usepackage{xcolor}
\newcommand{\edit}[1]{\textcolor{black}{#1}}

\newcommand{\expv}[1]{\ensuremath{\left\langle #1\right\rangle}}
\newcommand{\pairs}[2]{\left\langle #1\,#2\right\rangle}

\begin{document}

\title{Learning to find order in disorder}

\author{Humberto Munoz-Bauza}
\affiliation{Department of Physics and Astronomy, University of Southern California, Los Angeles, California 90089, USA}
\affiliation{Center for Quantum Information Science \& Technology, University of Southern California, Los Angeles, California 90089, USA}

\author{Firas Hamze}
\affiliation{D-Wave Systems, Inc. 3033 Beta Avenue, Burnaby, British Columbia, V5G 4M9, Canada}

\author{Helmut G. Katzgraber}
\affiliation{Microsoft Quantum, Microsoft, Redmond, Washington 98052, USA}
\affiliation{Department of Physics and Astronomy, Texas A\&M University, College Station, Texas 77843-4242, USA}
\affiliation{Santa Fe Institute, 1399 Hyde Park Road, Santa Fe, New Mexico 87501 USA}

\begin{abstract}

We introduce the use of neural networks as classifiers on classical
disordered systems with no spatial ordering. In this study, we \edit{propose a framework of design objectives for learning tasks on disordered systems. Based on our framework, we} implement
a convolutional neural network trained to identify the spin-glass state
in the three-dimensional Edwards-Anderson Ising spin-glass model from an
input of Monte Carlo sampled configurations at a given temperature. The
neural network is designed to be flexible with the input size and can
accurately perform inference over a small sample of the instances in the
test set.
\edit{We examine and discuss the use of the neural network in classifying instances from three-dimensional Edwards-Anderson Ising spin-glass in a (random) field}
%  Using the neural network to classify instances of the
%three-dimensional Edwards-Anderson Ising spin-glass in a (random) field
%{our results suggest an} inferred phase boundary is consistent with the absence
%of an Almeida-Thouless line.

\end{abstract}

\maketitle

\section{Introduction}

Machine learning methods are a class of artificial intelligence
algorithms that learn to perform tasks through the extraction of
patterns from data sets. Artificial neural networks, or simply neural
networks (NN), are machine learning methods inspired by biological
neural systems and are suitable for function approximation, image
classification, and various other pattern recognition tasks
\cite{haykin:08,goodfellow:16,bishop:06}.  Recently, machine learning
methods, and neural networks in particular, have also found applications
in computational condensed matter physics with \edit{phase transition
and complexity assessment in both classical and quantum systems
\cite{carrasquilla:17,chng:17,tanaka:17,kashiwa:18x},} as well as complex physical systems such as structural glasses \cite{ronhovde:11,nussinov:16}.  A potential advantage of neural
networks is their ability to generalize their learning when applied to a
different---but closely related---class of data. For instance,
Ref.~\cite{chng:17} used a convolutional neural network trained on the
Fermi-Hubbard model of correlated electrons at half-filled chemical
potential to infer the transition temperature away from half filling,
where the Hamiltonian suffers from the ``sign problem'' of quantum Monte
Carlo. This suggests that machine learning can provide insight in
situations where Monte Carlo methods may not be readily available or
might require exorbitant numerical effort.

Here we introduce the use of phase classifying machine learning methods
for spin glasses \cite{binder:86,mezard:87,young:98,stein:13},
archetypal disordered frustrated magnets for which there is little
theoretical understanding when the models are short ranged and where
numerical simulations are typically extremely difficult, thus requiring
vast amounts of CPU time. \edit{We design a convolutional neural network with the capacity to distinguish a spin-glass state from paramagnetic and ferromagnetic states in three dimensions at zero field. We introduce a set of design objectives as a proposed framework of necessary conditions for a machine learning task on spin glass microstates to be completed successfully, and we show that generalization is successful within this framework for spin glasses at zero field.}

 \edit{However,} possibly the most controversial aspect in the
theory of spin glasses is the existence of a spin-glass state in the
presence of a field.  While the replica-symmetry breaking picture of
Parisi~\cite{parisi:80} predicts a spin-glass state for short-range
systems, the ``droplet picture'' of Fisher and
Huse~\cite{fisher:86,fisher:88,moore:11,moore:12} states that any
infinitesimal (random) field destabilizes the spin-glass state at all
finite temperatures.  There have been numerous attempts to numerically
establish the existence of a spin-glass state in a field for short-range
systems with contradicting results. While some
\cite{caracciolo:90,houdayer:99,young:04,katzgraber:05c,sasaki:07b,joerg:08a,katzgraber:09b,larson:13}
find no evidence of a transition, other studies seem to detect a
spin-glass state in a field
\cite{marinari:98c,leuzzi:09,fernandez:10,baity:14,angelini:15}, i.e.,
the de Almeida-Thouless line \cite{almeida:78}.  In particular, there
has been disagreement on the different observables to be used
\cite{leuzzi:09,katzgraber:09b}, especially given the strong finite-size
effects typically observed in spin-glass simulations.  We note that the
possibility of a critical dimension above which a spin-glass state
occurs for a short-range system in a field has also been considered
\cite{newman:07}.

%To demonstrate the potential utility of machine learning and work around
%the issues imposed by different observables typically measured in
%spin-glass simulations, here we use a neural network to search for a
\edit{Understanding that the physics of spin glasses in a field is a subtle issue, 
we examine the potentially utility of machine learning tools to work around direct simulation observables and to attempt to find a}
stable spin-glass phase in the presence of a (random) field in the
three-dimensional Edwards-Anderson Ising spin glass\edit{. We primarily compare and contrast the machine learning predictions with the} 
results of Ref.~\cite{young:04}.

We implement \edit{our} neural network architecture using the TensorFlow
\cite{albadi:16} library for Python and find that it demonstrates strong
evidence of learning a representation of the three-dimensional
spin-glass state at zero field, if also given the opportunity to learn
from simple three-dimensional ferromagnetic data to additionally
distinguish a ferromagnetic state. The main advantages of our
implementation are  as follows. First, the approach makes classification
inferences on the basis of multiple instance samples of the spin-glass
Hamiltonian and multiple configuration samples from each instance.
Second, its input can consist of configurations of any linear system
size $L\geq 6$ by assuming periodic boundary conditions. We test the
ability of the NN to generalize its knowledge when the spin-glass
Hamiltonian has a nonzero field and find  the classification results to
be consistent with the Monte Carlo results of Ref.~\cite{young:04},
\edit{which finds through a correlation length analysis} that a spin-glass state is not stable in an external field for a
three-dimensional short-range system.
\edit{Due to the issues explained above, this is conclusion is limited by assumptions on the nature of the spin-glass state and the extent of finite size effects. We discuss potential directions using our proposal as a starting point that can address these assumptions.
}

The paper is structured as follows. In Sec.~\ref{sec:nn} we introduce
the model we study, followed by details of our NN implementation and
simulation details \edit{as well as a discussion of its limitations}. Results are presented in Sec.~\ref{sec:res},
followed by concluding remarks.

\section{Model, Computational Methods and Neural Networks}
\label{sec:nn}

\subsection{Model}

The Edwards-Anderson Ising spin glass model \cite{edwards:75} is described
by the Hamiltonian
\begin{equation}
H=-\sum_{\pairs{i}{j}} J_{ij} s_i s_j - \sum_i h_i s_i,
\end{equation}
where each $J_{ij}$ and $h_i$ is a random variable drawn from a given
symmetric probability distributions. \edit{We refer to each such sample as an \emph{instance} of the Ising spin glass model, while each microstate of spin configurations is called a \emph{sample} of the instance.} In zero-field models, $h_i = 0$
$\forall i$.  In this paper we primarily consider the couplings $J_{ij}$
to be drawn from a Gaussian distribution of unit variance. In nonzero
field models, each field $h_i$ is drawn from a Gaussian distribution
standard deviation $h$.

\subsection{\edit{Design framework} of our neural network}
\label{sec:des}

Carrasquilla and Melko \cite{carrasquilla:17} implemented a multi-layer
perceptron (MLP)---a neural network with a single hidden layer---to
distinguish the ferromagnetic and paramagnetic states on the
two-dimensional Ising model and detecting its phase transition regime by
finding the point where the classification probabilities cross. Using
arrays of spin configurations sampled from Monte Carlo simulations as
inputs, their neural network was trained to classify single
configuration samples as being in a ferromagnetic or paramagnetic state.
With the resulting classification probabilities, one can identify a
transition temperature $T_c$ from where the probabilities cross. The
neural network successfully discriminates between the ferromagnetic (FM)
and paramagnetic (PM) phases for linear system sizes between $L=10$ and
$L=60$, with a natural classification error occurring in the vicinity of
the phase transition temperature $T_c \approx 2.269$ \cite{yeomans:92}.
They find that the representation of the ferromagnetic phase in the MLP
is tied to learning the average magnetization $m$ by comparing it with a
toy-model MLP in which the hidden layer is simplified to directly
measure $m$, while leaving a single free parameter that is learned
during training.

Following the initial results of Ref.~\cite{carrasquilla:17}, Ch'ng {\em
et al.}~\cite{chng:17} implemented a three-dimensional convolutional
neural network (CNN) to distinguish the N\'eel transition in the
three-dimensional Fermi-Hubbard model at half-filling, and extrapolated
its magnetic phase diagram as the chemical potential is varied, which is
problematic to study as the sign problem of quantum Monte Carlo becomes
significant.  This paper has the analogous objective of identifying the
spin-glass state in the Edwards-Anderson model and establishing its
phase boundary, with the benefit that past Monte Carlo studies are
available to compare with.

An outline of the design framework for a phase classifying neural network on spin glasses is as follows:
\begin{enumerate}

\item The spin configuration \edit{samples} of the spin-glass model at low
temperatures are visually indistinguishable from paramagnetic ones. To
overcome this problem, as inputs for learning we use replica
overlap \edit{samples}
\begin{equation}
q_i = s_i^{(1)}s_i^{(2)}
\end{equation}
taken from two independent \edit{spin samples} $(1)$ and $(2)$. The usual
overlap parameter $q$ \cite{binder:86} is the \edit{mean} of the $q_i$ with 
$i \in \{1, \ldots, N\}$ and $N = L^3$ the number of spins.

\item The networks in Ref.~\cite{carrasquilla:17} infer a phase using a
single Monte Carlo configuration sample at a time as the input to the
neural network. To ensure reliable phase classification, we implement a
CNN that averages over multiple configuration samples from a Monte Carlo
simulation.

\item To learn a reasonable representation for the spin-glass state, a
neural network should consider configurations across multiple instances
simulated at the same temperature $T$ and with the same field strength
$h$, as the critical properties for a spin-glass transition are
considered from a quenched average. Thus, we propose that an averaging
step over a sample of instances (instance averaging) is necessary for a
NN to classify the spin-glass state faithfully.

\item It is possible that a supervised binary classifier for the spin
glass (SG) and PM phases may learn to represent the SG state in a way
that is only sensitive to the magnitude of the overlap parameter $q$.
While such a classifier would probably be successful in a zero-field
model, one could question whether such a classifier has truly
``learned'' the spin-glass state. However, teaching a classifier to also
distinguish between the SG and FM phases helps create a better knowledge
representation of the SG state that is not too directly tied to $|q|$.
For both a typical spin-glass replica and for the Ising ferromagnet,
there is a large probability of $q\approx\pm 1$ below the critical
point. However, intermediate values of $q$ between $-1$ and $1$ are
likely only in a spin-glass model due to nontrivial ergodicity breaking.
Thus, this motivates a design for a phase classifier that is taught {\em
three categories}, i.e., a ternary classifier, of the FM, PM, and SG
states.

\item When generalizing to the nonzero field case, the mean of each
$q_i$ is biased because the spins become more likely to point in the
direction of their local field. Thus, each $q_i$ should be centered
around its thermal mean $\expv{q_i}$.

\end{enumerate}

\subsection{\edit{Limitations of the framework}}

\edit{
While our framework sets the conditions to make the machine learning of 
the spin-glass state reachable, 
it is not without likely limitations in applicability and explanatory power. 
It establishes the machine learning task as that of one where a neural network learns the typical pattern of spatial correlations 
in thermally sampled spin overlap configurations from typical instances of spin glasses.
In the non-zero field case of spin glasses, atypical values of $q$ during Monte Carlo simulations can cause strong fluctuations that may lead to an unsuccessful finite size scaling analysis \cite{baity:14a, parisi:12}. 
Likewise, one could expect that the dominance of rare events in $q$
may reduce the capacity for a neural network to learn and identify a phase in the thermodynamic limit.}

\edit{
Suppressing rare-event fluctuations and constructing a processed data set that is more representative of the thermodynamic limit would likely 
improve the performance of machine learning methods by reducing the amount of noise in the data. 
While we don't apply additional processing on our data set here, 
such a method could be based on simply constraining the sampled value 
of $q$ to a fixed value and training on the filtered data set.
}

\edit{
While we have proposed the generalization task of classifying a phase at non-zero field based on training at zero-field, 
there is substantial theoretical and numerical evidence that the spin-glass state is inherently different at non-zero field
\cite{leuzzi:09, young:98}. As such, simply training on FM, PM, and SG states at zero field may result in a learned representation that is not accurate in the thermodynamic limit at non-zero field.
}

\edit{
In spite of theoretical concerns for three dimensional systems, our proposed framework is not necessarily restricted to 3D spin glasses. 
One could in principle implement analogous higher-dimensional 
convolutions to probe the physics of higher-dimensional spin glasses 
closer to the critical dimension. The four-dimensional case would be a 
natural modification, given its more numerically conclusive AT line \cite{banos:12}. 
However, this is not without some additional software design effort, 
as high $N$-d convolutions are neither ubiquitous in the machine learning field 
nor natively supported in the Tensorflow API (see \url{https://github.com/tensorflow/tensorflow/issues/9513}).
}

\subsection{Convolutional neural network and training}

\begin{figure*}
\includegraphics[width=2.0 \columnwidth]{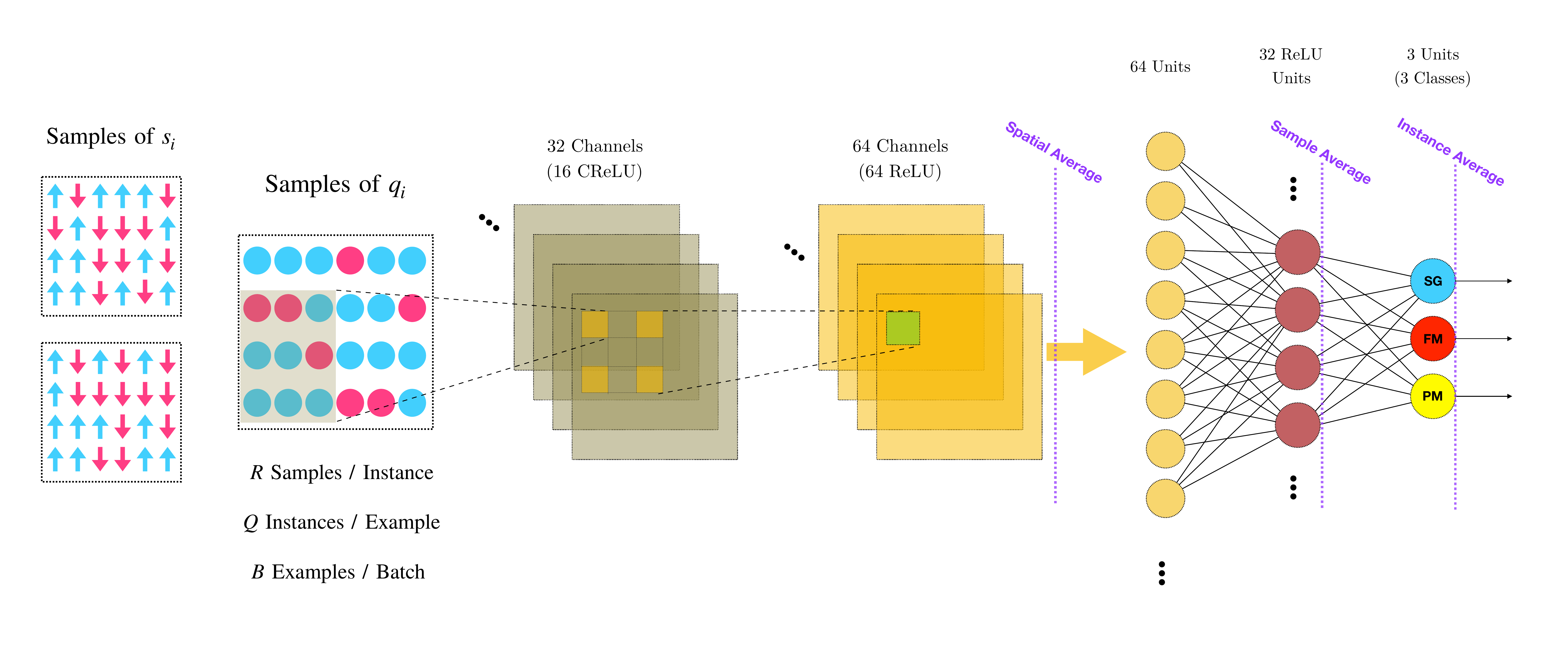}
\caption{\label{fig:cnn} Illustration of the convolutional neural network evaluation procedure outlined in Algorithm \ref{alg:sgnet}. Two-dimensional images are depicted for clarity, but the input and convolutional layers are three-dimensional.
}
\end{figure*}

The ternary state classifying neural network (hereafter referred to as
the {\em classifier}) is a three-dimensional convolutional neural network
with two convolutional layers that extract spatial information from the
overlap samples from each instance, followed by two fully-connected
layers with intermediate averaging steps. The classifier is illustrated in Figure \ref{fig:cnn} and detailed in
Algorithm \ref{alg:sgnet}. The predictions of the CNN for a single
example manifest in line 10 as a vector $y_i$ of three real numbers,
which are not necessarily bounded. The last step applies the ``softmax''
function to create probabilities
\begin{equation}
p_i = \frac{e^{y_i}}{\sum_j e^{y_j}}
\end{equation}
that can be interpreted as the degree of belief that a given example
belongs to in each class (state) $i$. We refer to these probabilities as
{\em softmax probabilities}. However, the results presented here utilize
a {\em classification probability} of a category $i$, which is the rate
that $i$ is the most likely class (largest $p_i$) for the instances at a
given temperature. This is the more practical measure for testing a NN.

During training we wish to align the prediction of the CNN as close as
possible to the actual label of the example by adjusting the parameters
of each layer of the CNN. In other words, we would like to match the
softmax probabilities $p_i$ as close as possible with the indicator
probabilities $d_i$, where $d_i=1$ if $i$ is the index of the correct
state, and $0$ otherwise. This can be cast as a numerical optimization
problem of a cost function $C(p_i,d_i)$. Here, the cost function is the
negative cross entropy of the probabilities,
\begin{equation}
C(p_i,d_i) = -\sum_i\left[d_i\ln p_i + (1-d_i)\ln(1-p_i)\right].
\end{equation}
The general strategy in machine learning is to calculate the numerical
gradient of the cost function for a batch of examples and adjust the NN
parameters in the negative gradient direction until a minimum for $C$ is
reached. The simplest gradient strategy is stochastic gradient
descent (SGD)\edit{, where each iteration of optimization uses a  random batch of examples.} However, in this paper we use the ADAM \cite{kingma:14}
method, an adaptive variant of SGD, for training the parameters of the
neural network along with an $\ell^2$ penalty on the cost function. Further
details on NN training can be found, for example, in 
Refs.~\cite{haykin:08} and \cite{goodfellow:16}.

In Algorithm \ref{alg:sgnet}, the hyperparameters $B$, $Q$, and $R$
specify the grouping sizes of the training examples for every training
step. $B$ is the number of examples in each step (the batch size), $Q$
is the number of spin-glass instances randomly selected in each example,
and $R$ is number of overlap configuration samples for each instance.
Per the second and third \edit{point of the design framework} in Sec.~\ref{sec:des}, a single
example is a $2$-tensor of three-dimensional configurations, with
dimensions $Q \times R \times L^3$, which is progressively reduced
through the layers of the CNN into the vector $p_i$ of softmax
probabilities. In every step of training, a batch of these examples of
size $B$ is evaluated into softmax probabilities, and the cost function
is averaged across the batch to evaluate the gradient of the cost
function.

The convolutional layers are ``pool-less,'' i.e., the output tensor
sizes are instead controlled through the use of kernel strides and
periodic padding of the overlap samples.
The primary activation function used is the rectified
linear unit (ReLU) function, which simply applies $\max(x,0)$ on each
element $x$ of the outgoing tensor. The first convolutional layer uses
the concatenated ReLU (CReLU) variant activation function, which uses
twice the number of output channels as kernels to additionally utilize
the negatives of learned features. This avoids the neural network having
to duplicate learning effort on closely related kernels \cite{shang:16}.

{The first layer is capable of capturing correlations within a three
spin radius. As the kernels have a stride of two, information redundancy
is minimized while not completely decimating the spins into blocks of
side length three. In the second layer, dilation \cite{yu:15} introduces
the capacity for the CNN to represent longer-range correlations. In our
model, each individual unit in the second convolutional layer is capable
of capturing correlations within a region with a side length of $7$ spins.
Thus, if a large number of units in a particular channel capture the
correlations they are trained to find, the spatial average is capable of
expressing any long-range correlations that may be present. While deep
neural networks excel at integrating such correlations for inference
task, it should be advised for our particular task that they may not
necessarily make use of features that are directly physically relevant
after training. However, in order to successfully distinguish the SG
from the FM and PM states, the neural network should at minimum be
able to recognize nontrivial wave-vectors of the spin overlap, hence our
inclusion of all three states in our training procedure, and our choice
of basing the neural network around two convolutional layers, each with
a large receptive field.}

{The systematic testing of models with additional layers and parameters
would substantially increase the computational time and complexity of
our intended objectives. Thus, the design we present here is only one
possible example with the desired learning capacities, as discussed
above. Adding layers and fine-tuning the model would likely increase the
test accuracy, but at diminishing returns for the mere task of
classifying between three possibilities. }

\begin{figure}
\begin{algorithm}[H]
\caption{Convolutional neural network design\newline 
\hspace*{6.75em}for state classification on cubic lattices}\label{alg:sgnet}
\begin{algorithmic}[1]
\Procedure{SG-NET-EVAL$\;$}{Constants: $B$ batch size, $Q$ instances per example, 
$R$ overlap samples per instance; 
Input: $B\times Q\times R \times L^3$ tensor $T$ of overlap units.}
\State Center overlap samples around the replica mean (dimension $R$).
\State Pad all three spatial dimensions with 
$4-[L-1 \pmod 4]\pmod 4$ two-dimensional slices from each opposite end 
(Periodic padding ensures the input spatial length $L'$ is $1 \mod 4$).
\State \textbf{Convolution 1} --- $16$ CReLU activated channels 
(effectively $32$ output channels). Kernel length $3$. Kernel stride $2$. 
Output spatial length $(L'-1)/2$.
\State \textbf{Convolution 2} --- $64$ ReLU activated channels. Kernel 
length $2$. Dilation length $1$. Output spatial length $(L'-1)/4$.
\State Reduce mean over all spatial dimensions. Output tensor size
$B\times Q\times R\times 64$.
\State \textbf{Fully-connected layer 1} --- $32$ ReLU activated units applied on 
the last dimension. Output tensor size $B\times Q\times R\times 32$.
\State Reduce mean over the overlap sample dimension. Output tensor size 
$B\times Q\times 32$.
\State \textbf{Fully-connected layer 2} --- $3$ linear units applied on the last 
dimension. No bias or activation. Output tensor size $B\times Q\times 3$.
\State Reduce mean over the instance dimension. Output tensor size $B\times 3$.
\State Softmax and inference across the batch, where $0$ is the PM state,
$1$ is the FM state, and $2$ is the SG state.
\EndProcedure
\end{algorithmic}
\end{algorithm}
\end{figure}

\subsection{Monte Carlo training sets}

We use parallel tempering (PT) Monte Carlo \cite{hukushima:96} to
generate \edit{spin} configurations from spin-glass instances, as well as a few
ferromagnet runs.  Sample decorrelation is enhanced by two concurrent,
independent runs of PT for each instance. Four concurrent runs are used
in the case of nonzero field spin glasses. {Our training procedure
involves the shuffling and sampling of spin states obtained during Monte
Carlo sampling. In combination with PT and training in batches of
multiple instances, the trained neural network is expectedly robust
against time correlations.}

\begin{table}
\caption{Monte Carlo parameters for the instances used for training, as
well as $100$ instances set aside as a validation set for the confusion
method. $N$ is the number of variables,  $N_{\text{sw}}$ is the number
of Monte Carlo sweeps, $N_{\rm s}$ is the number of configurations taken
times independent PT runs and $N_T$ is the number of temperatures used.
EA represents the Edwards-Anderson model, FM the Ising ferromagnet.
}\label{tbl:tparams}
\begin{ruledtabular}
\begin{tabular}{lcccccr}
Type & $L$ & $N$ &  $\log_2 N_{\text{sw}}$ & $N_{\rm s}$ & $N_T$ & $[T_{\rm min},T_{\rm max}]$\\
\hline
EA &  8 & 3000 & 22 & $32\times 2$   & 20 &  [0.20, 2.00]\\
EA & 12 & 600  & 25 & $32\times 2$   & 20 &  [0.20, 2.00]\\
EA & 12 & 100  & 25 & $32\times 2$   & 20 &  [0.20, 2.00]\\
FM &  8 & 1    & 24 & $4096\times 4$ & 32 &  [2.00, 7.00]\\
\end{tabular}
\end{ruledtabular}
\end{table}

\begin{figure}
\includegraphics[width=\columnwidth]{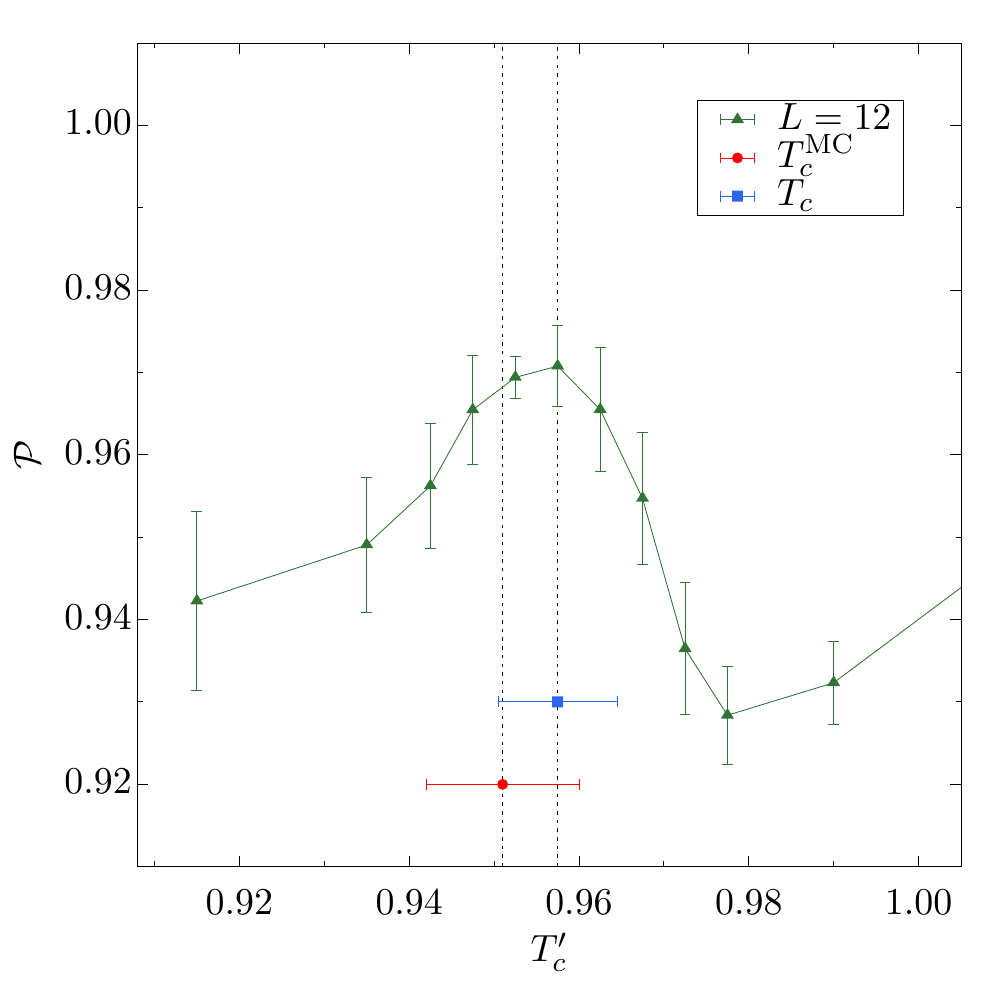}
\caption{\label{fig:conf}
Classifier accuracy as measured by Eq.~\eqref{eq:acc} as the transition
temperature in training is varied. An estimate of the inferred $T_c$
(blue square) and the Monte Carlo value found in Ref.
\cite{katzgraber:06} (red circle) are also depicted.}
\end{figure}

We generate instances for the training data as outlined in Table
\ref{tbl:tparams}. The system size of $L=8$ is small enough to
thermalize easily and thus to simulate a large number of instances,
yet large enough to fit with larger systems in a finite-size scaling
analysis, e.g., as done in in Ref.~\cite{katzgraber:06}. A large amount
of $L=8$ samples can thus provide the classifier a ``big picture" view
of the difference between the spin-glass state and the paramagnetic
state with much more uncorrelated data. A larger system size ($L=12$) is
then useful for improving the classifier's characterization near the
critical temperature. For our implementation, the size parameters of
Algorithm \ref{alg:sgnet}, the learn rate parameter $\eta$, and the
$\ell^2$ penalty weight parameter $\lambda$, are specified in Table
\ref{tbl:hparams}. Error bars in classification probabilities and
performance are derived from independently training the neural network
$16$ times on the training instances.

Because the classifier relies on supervised training, a critical
temperature needs to be presumed before labeling the data. While this
could be estimated with Monte Carlo statistics or quoted from previous
results, we instead utilize the confusion method introduced in 
Ref.~\cite{nieuwenburg:17} to find the critical temperature. We repeat the
entire training procedure for various choices of $T'_c$ within the
critical region, and select the choice that maximizes the accuracy of
the classifier. We choose as our accuracy measure:
\begin{IEEEeqnarray}{rCl}
\mathcal{P} &=& \int_{T_{\rm min}}^{T_{\rm max}}\text{d} T\,\,
	\left[ p_{\rm SG}(T)\theta(T'_c -T)\right. \IEEEnonumber \\ 
	&&\qquad \qquad\quad \left. +\, p_{\rm PM}(T)\theta(T-T'_c)\right],
\label{eq:acc}
\end{IEEEeqnarray}
where $p_{\rm SG}$ and $p_{\rm PM}$ are the classification probability
\emph{densities} of the SG and PM states between $T_{\rm min}$ and
$T_{\rm max}$, respectively. Naturally, because data are only collected
for a discrete set of temperatures, ${\mathcal P}$ should be calculated
by a trapezoid-rule summation of the classification probabilities
divided by the temperature interval measure $T_{\rm min}-T_{\rm max}$.
It is important to weigh the classification probabilities as a numerical
integral--and not just sum them directly--to take the uneven spacing of
the simulation temperatures into account.

\begin{figure*}
\includegraphics[width=2.0\columnwidth]{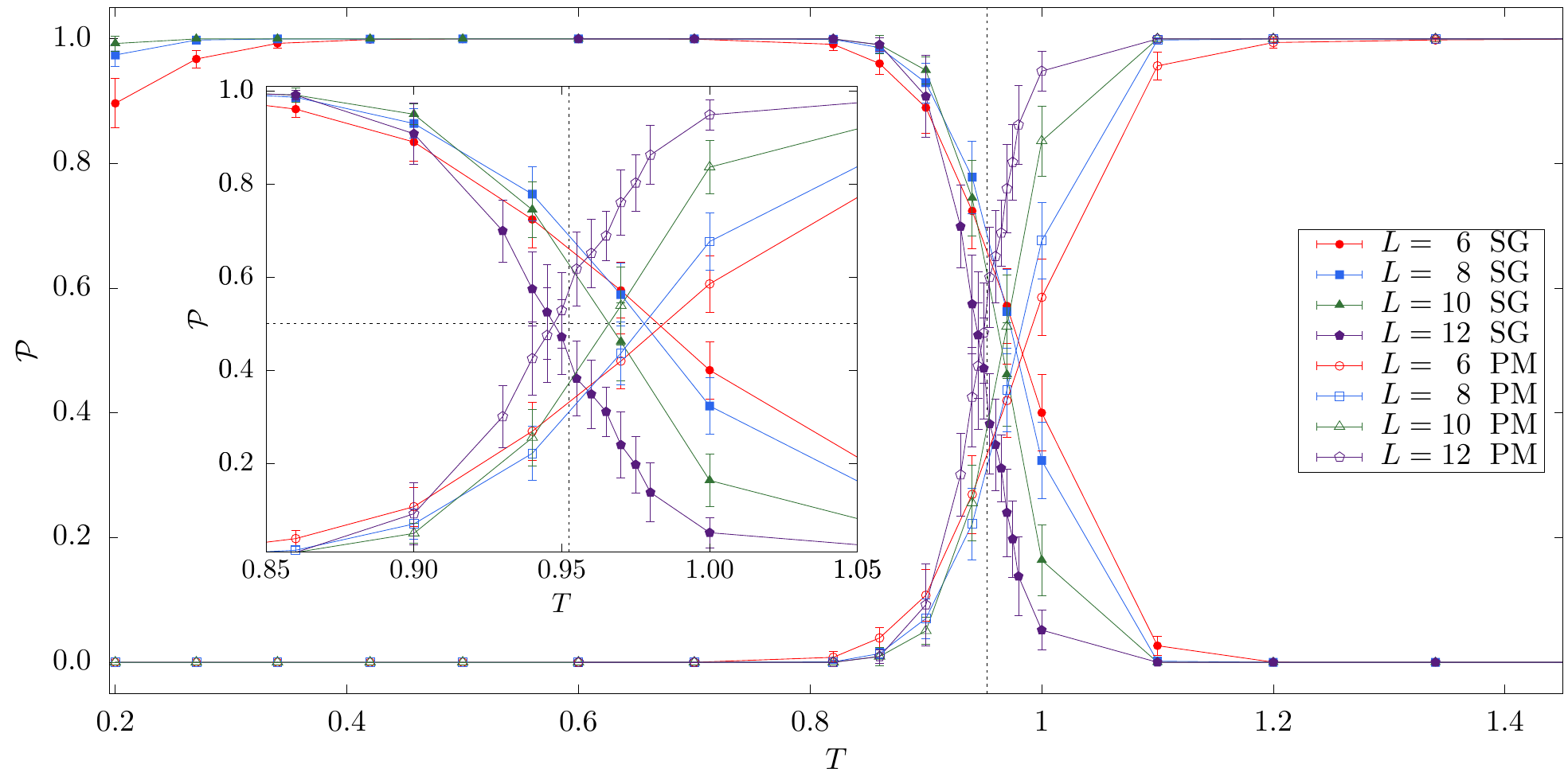}
\caption{\label{fig:evalsg1}
Classification probabilities of the CNN for finding a spin glass (SG) or
paramagnetic (PM) state. Where the probabilities do not add up to $1$,
the remainder belongs to FM classification probability, which is omitted
for clarity. The figure inset zooms into the region between $T=0.85$ and
$T=1.05$ where the transition in classification probability occurs for 
better visibility.}
\end{figure*}

To simplify training, the critical temperature for classifying the FM
state is simply quoted as $T_c=4.51$ \cite{binder:01}. We do not attempt
to optimize this transition temperature with the confusion method.

\begin{table}
\caption{Training parameters for the neural network. Training is divided
into three stages where the parameters are adjusted for the specified
amount of steps. $\eta$ represents the learn rate parameter, $\lambda$
the penalty weight parameter, $B$ is the number of examples, and $Q$
and $R$ specify the grouping sizes for the training examples for each
training step.
}
\label{tbl:hparams}
\begin{ruledtabular}
\begin{tabular}{lccccr}
 Steps & $\eta$ & $\lambda$ & $B$ & $Q$ & $R$\\
\hline
2000 & $10^{-3}$ & $0.005$ & 12 & 6 & 16\\
4000 & $10^{-4}$ & $0.005$ & 12 & 6 & 16\\
4000 & $10^{-5}$ & $0$     & 12 & 6 & 16
\end{tabular}
\end{ruledtabular}
\end{table}

\section{Results}
\label{sec:res}

\begin{table}
\caption{Table of parameters for each instance class simulated for a
given linear size $L$ and field strength $h$ for the test set of
examples.  See Table \ref{tbl:tparams} for additional details.}
\label{tbl:params}
\begin{ruledtabular}
\begin{tabular}{clcccc}
$L$ & $h$ & $N$&$\log_2 N_{\text{sw}}$&$ N_{s}$ & $N_T$\\
\hline
6  & 0 & 1000 & 23 & 64  & 20\\ 
8  & 0 &  400 & 23 & 128 & 20\\
10 & 0 &  400 & 24 & 64  & 20\\
12 & 0 &  100 & 25 & 64  & 20\\
\hline
10 & 0.025 & 500 & 24 & 64 & 16\\
6  & 0.05  & 500 & 23 & 64 & 16\\
8  & 0.05  & 500 & 23 & 64 & 16\\
10 & 0.05  & 500 & 24 & 64 & 16\\
10 & 0.075 & 500 & 24 & 64 & 16\\
6  & 0.10  & 500 & 24 & 64 & 16\\
8  & 0.10  & 500 & 24 & 64 & 16\\
10 & 0.10  & 500 & 24 & 64 & 16
\end{tabular}
\end{ruledtabular}
\end{table}

\subsection{Confusion ensemble}

The confusion method is carried out with a validation set of $100$
spin-glass instances with $L=12$, which were not used for training the
classifier. The average accuracy as a function of $T'_c$ shown in
Fig.~\ref{fig:conf}.  The accuracy peak suggests a transition
temperature of approximately $T_c=0.958(8)$. This is consistent with the
critical temperature of the Edwards-Anderson model with Gaussian
disorder as found, for example, in Ref.~\cite{katzgraber:06}.  The
precise temperatures used as the boundary between the PM and SG states
that are within the range of both $T_c$ values are $0.953$ and $0.958$.
In the figures that follow, we use the classifier model trained at $T'_c
= 0.953$, however, the results from either model are practically
identical. 

The configuration samples for testing the classifier are
obtained with parallel tempering Monte Carlo simulations,
\edit{ with parameters listed in Table \ref{tbl:params}. For spin glasses in a field, we use the same temperature range and at least 10 times the number of sweeps as Ref.~\cite{young:04} used up to $L=8$ to ensure termalization. The samples are collected after a thermalization period of 75\% of the sweeps, for a time between samples of $2^{16}$ sweeps. }

\subsection{Spin glasses at zero field}

Figure \ref{fig:evalsg1} shows the classification probabilities of the
classifier evaluated on the sample of spin-glass test instances.  In all
figures, vertical dashed lines represent transition temperatures and
horizontal dashed lines the 50\% probability line.  The classification
probabilities follow a smooth transition near the critical temperature.
The classifier performs well, except for an anomalous jump in FM
classification at the lowest temperature. This is likely due to the
possibility of sampling instances with simple energy landscapes when
dealing with finite-size spin glasses. This is supported by observing
the finite size behavior as $L$ increases to $12$, for which the FM
anomaly vanishes.

\subsection{Ferromagnetic classification probabilities}

\begin{figure}
\includegraphics[width=\columnwidth]{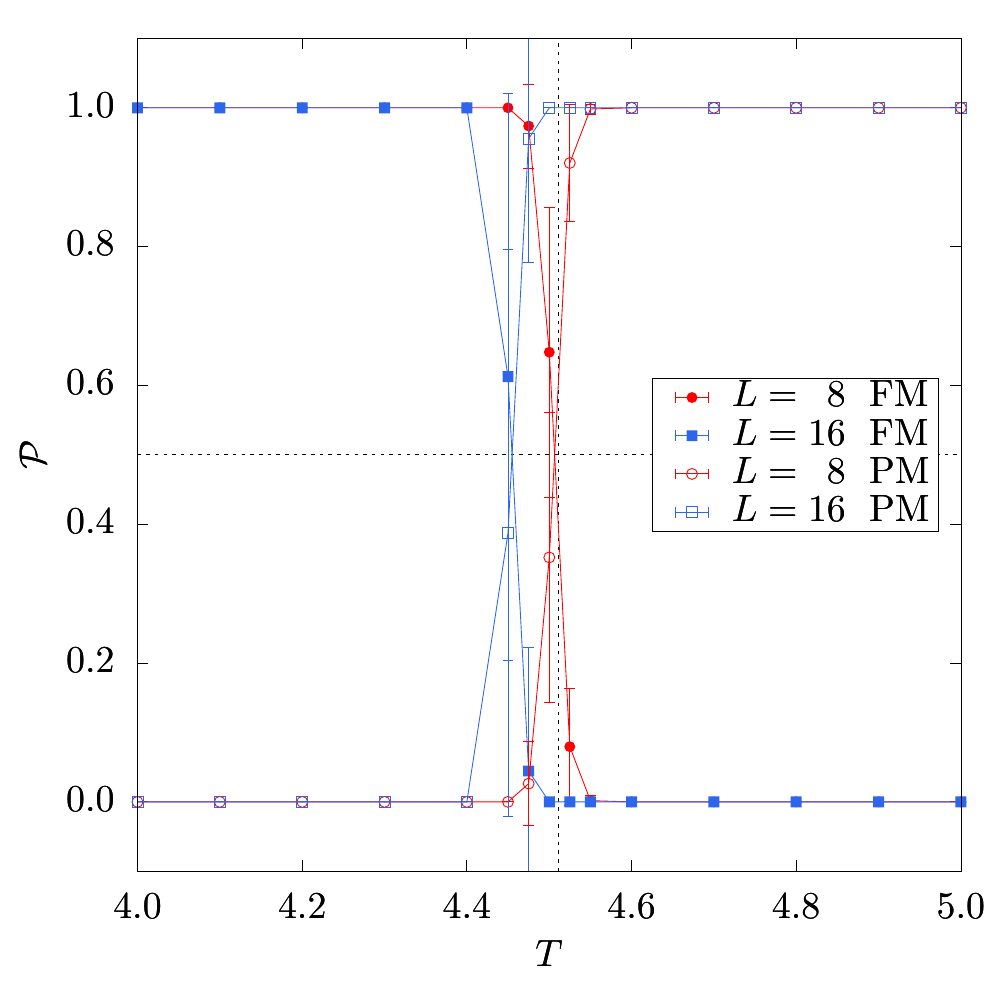}
\caption{Classification probabilities for the three-dimensional Ising
ferromagnet. Although finite-size effects are visible, these data 
are only used to illustrate the detection of the FM state.
}\label{fig:fm}
\end{figure}

Figure \ref{fig:fm} shows the classification probabilities of the
classifier on a test set of three-dimensional Ising ferromagnet samples
Note that in the CNN algorithm, the instance averaging step is kept for
code consistency. An ``instance'' of the Ising ferromagnet is simply
another set of uncorrelated configuration samples. The distinction that
the classifier learns is affected by finite-size effects, i.e., the
labeling that the classifier learns for $L=8$ ($T'_c=4.51$) shifts for
$L=16$ ($T'_c =  4.46$) to a lower value. \edit{Observables that would diverge at the phase transition, such as heat capacity, 
in the thermodynamic limit are instead critical at higher temperatures for small system sizes. 
The shift away from $T_c$ at system sizes larger the training set is expected 
since the larger system sizes become critical at slight lower temperature than the smallest system size. 
This causes the predicted transition of a larger system to seem lower than
the learned representation.} 

The classifier's sharp
transition region is likely a consequence of the number of
available training parameters in the model, compared to the complexity
of the learning task of distinguishing between the PM and FM states on a
given system size, as well as the enhancement furnished by averaging
over multiple configuration samples.

In principle, the characteristics of these classification probabilities
may be useful as a NN-based finite-size scaling technique. However,
neither learning the three-dimensional Ising ferromagnetic transition
temperature to high precision nor inferring its critical properties from
the NN were initial objectives in this work.

\subsection{Generalization to bimodal disorder}

\begin{figure}\label{fig:bm}
\includegraphics[width=\columnwidth]{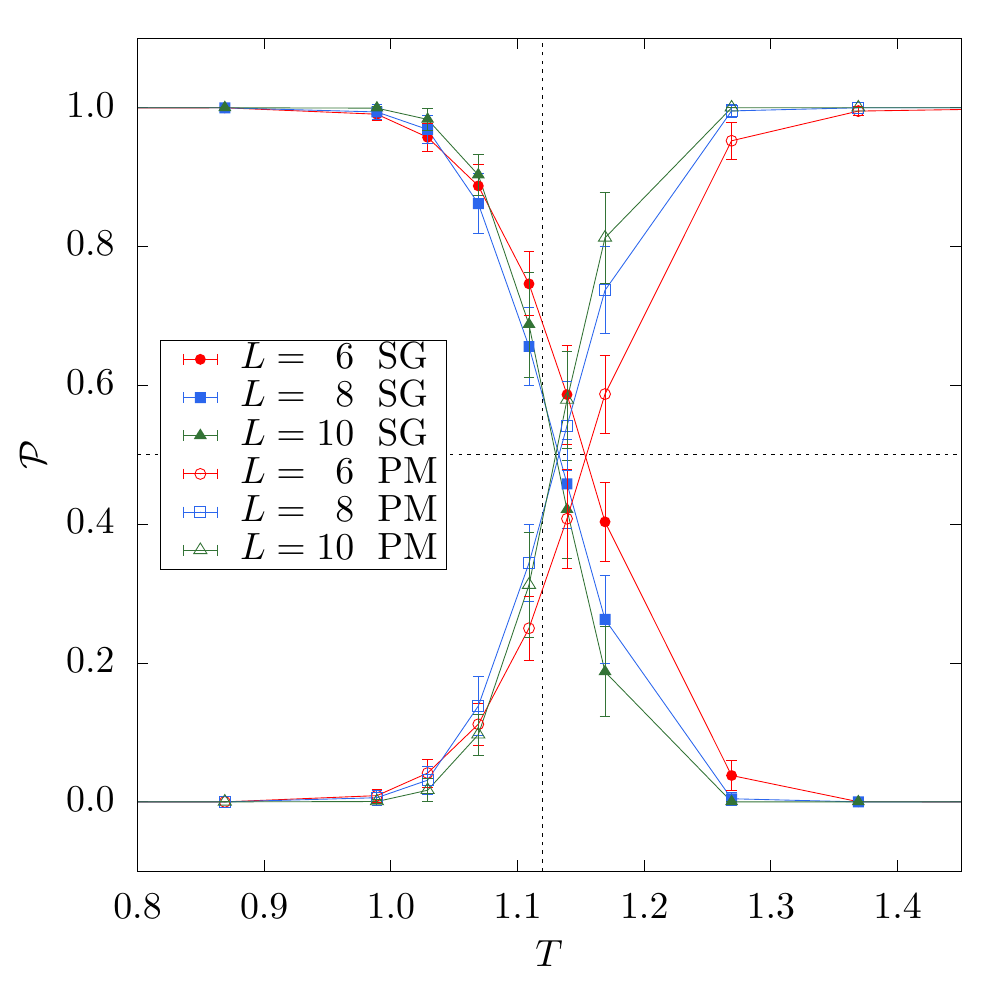}
\caption{
Classification probabilities of CNN trained on the Edwards-Anderson spin
glass with Gaussian disorder with a test set of instances with bimodal
disorder. The data cross close to $T'_c \approx 1.1$, the known value of
the bimodal spin-glass transition temperature.}
\end{figure}

As a first generalization test, we generate a test set of spin-glass
instances with bimodal instead of Gaussian disorder.  In this case, the
critical temperature shifts to $T_{c} \approx 1.12$
\cite{katzgraber:06}. The crossover point easily shifts to this new
temperature. We take this as very good evidence that the classifier
learns a good representation of the spin-glass state, and can make
accurate predictions for different spin-glass models with different
disorder distributions.

\subsection{Generalization to nonzero fields}

\begin{figure}
\includegraphics[width=\columnwidth]{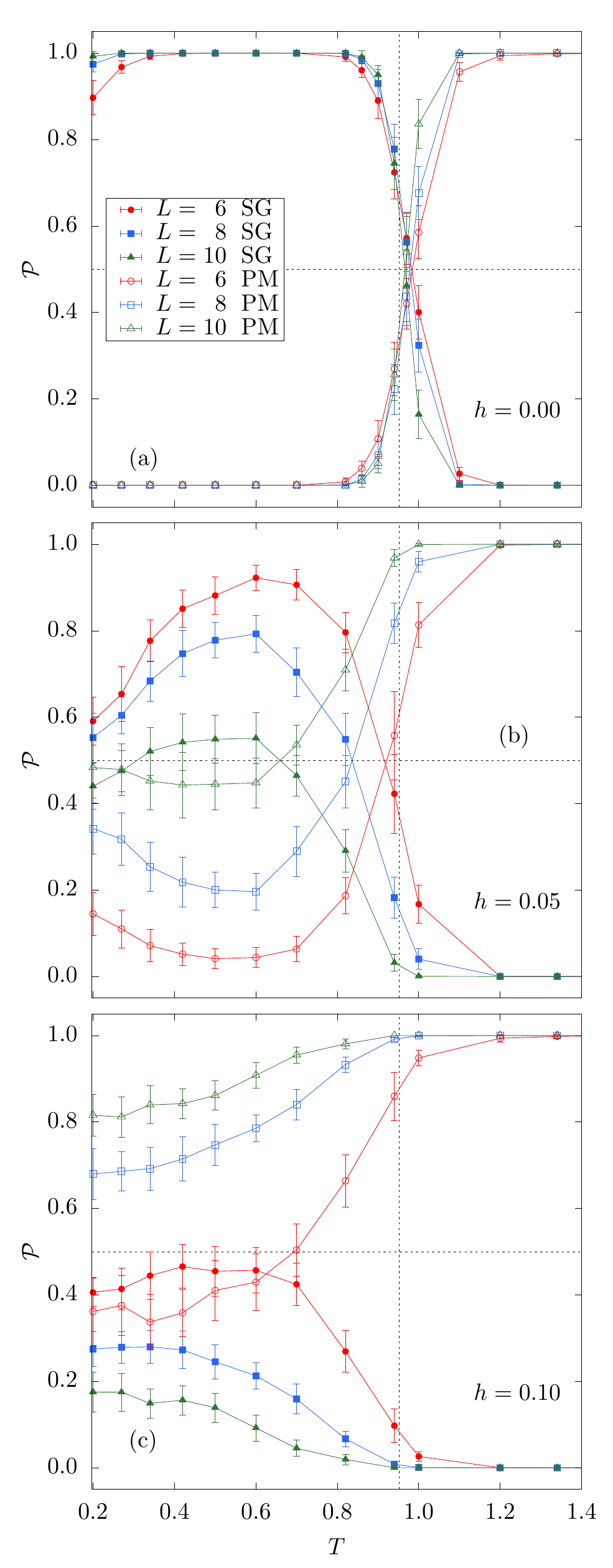}
\caption{Progression of classification probabilities for different
fields. As the field increases, the SG signal weakens progressively.
Even for fields as weak as $h = 0.05$ the signal of the transition
moves to zero temperature as the system size increases.
Data for (a) $h = 0.00$, (b) $h = 0.05$,  and (c) $h = 0.10$.
}
\label{fig:all_fields}
\end{figure}

\begin{figure}
\includegraphics[width=\columnwidth]{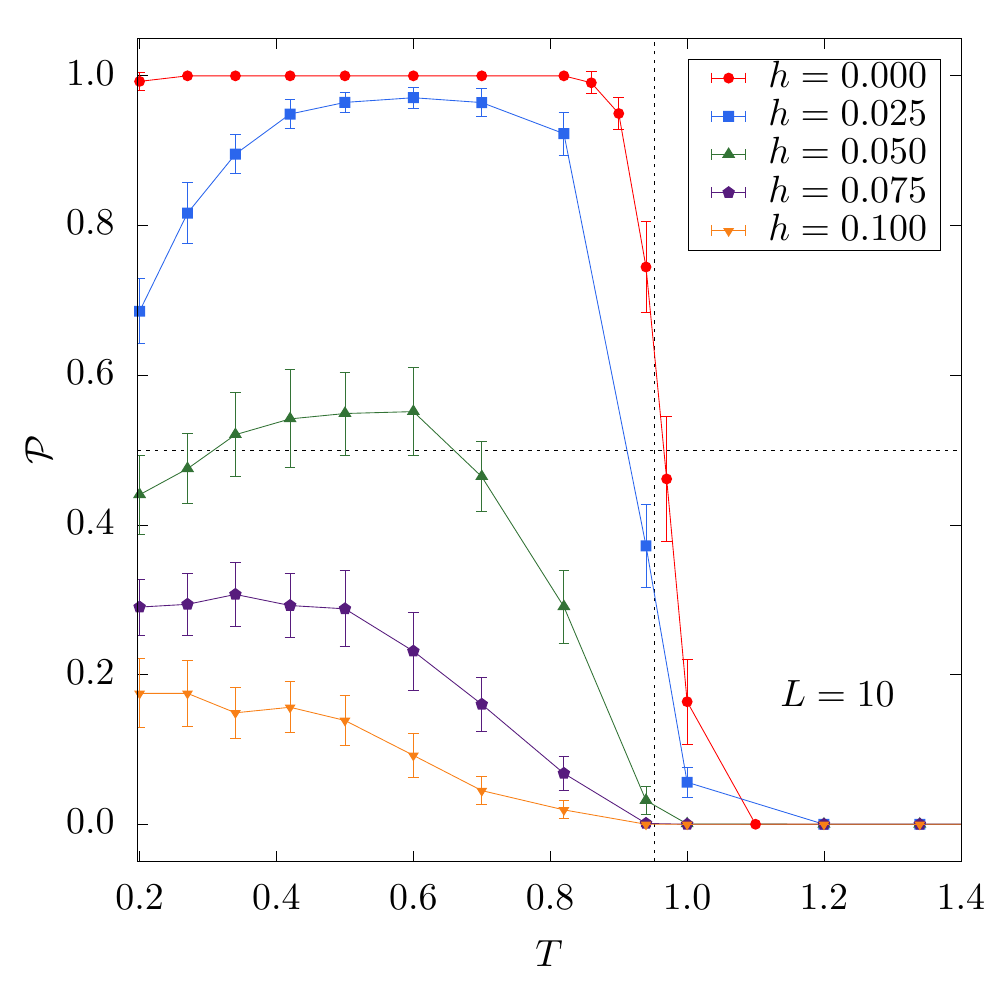}
\caption{Progression of classification probabilities with increasing
\edit{field} for $L = 10$. The stronger the field, the weaker the signal
for a SG state.}
\label{fig:each_size}
\end{figure}

Now that we have demonstrated that our NN can detect a spin-glass
transition at zero field and even detect correct transition temperatures
when the disorder is changed, we \edit{evaluate its prediction for} a spin-glass
state in a field.  The classification probabilities for nonzero fields
are shown in Figure \ref{fig:all_fields}. It can be observed that 
at $h=0.05$, the spin-glass signal for the classifier becomes weak.  At
a stronger field, $h=0.10$, the PM phase becomes dominant with larger
linear system sizes $L$. Figure \ref{fig:each_size} illustrates
a break down of any crossing temperature beyond system sizes of $L=6$.

\edit{
We emphasize again that the model is trained under the limitations explained for the framework,
and the system sizes examined here may not be conclusive in the thermodynamic limit. However, at least for the system sizes examined here, we can observe that the transition from the spin-glass state 
to the PM state with increasing $h$ becomes sharper as the system size is increased.
From this trend, one would conclude from the results of the classifier that if there is a spin-glass state in a field, it occurs for fields $h \lesssim 0.05$. This is consistent with the correlation length-based analysis of Ref.~\cite{young:04} for similar system sizes.
}

%A possible objection could be that, for $L=10$, there appears to be a
%region between $h=0.0$ and $h=0.05$ where the classification probability
%is significant. Could a transition simply be somewhere there? As was
%pointed out, e.g., in Ref.~\cite{young:04}, fields of strength $h
%\approx 0.1$ are far below the predicted critical field from mean-field
%calculations, namely $h_{\rm AT} \approx 0.65$. The results presented here
%thus suggest that if there is a spin-glass state in a field, it must occur for
%fields $h \lesssim 0.05$.

\section{Summary and conclusion}

We have implemented and evaluated an input-size agnostic
three-dimensional convolutional neural network for ternary phase
classification on Ising-like models three-dimensional cubic lattice and
{found evidence that such a neural network can acquire} an accurate
representation of the spin-glass state {at zero field}.  We emphasize
that {performing this classification task} is nontrivial, because of the
lack of spatial magnetic order in these systems. Furthermore, the
approach can be generalized to other three-dimensional models.

Conditioned on training at zero field only, the classification inference
of the neural network at nonzero field \edit{would suggest} that a spin-glass state
in a field might not be stable for three-dimensional systems. This
inference comes with \edit{the caveats discussed above.}
%First, the neural network
%inference is performed without the use of standard order parameters, and
\edit{Most significantly} there is no \emph{a priori} theoretical reason to expect the neural
network's accuracy on data sets of a significantly different nature to be
high. As such, phase-classifying neural networks are by no means
intended to replace or directly compare with Monte Carlo sampling of
observables, and have utility instead as a cost-effective, qualitative
guide for exploring the phase space of condensed matter systems outside
of well-studied regimes. \edit{We} emphasize that the numerical effort
needed \edit{for the learning and classification tasks here can be made considerably small}: Whereas most spin-glass
studies require tens of thousands of samples {at nonzero field}, \edit{with the neural network} {a likely conclusion} can be drawn from as little as $500$ disorder
instances, and at minimum helps to \edit{identify} the relevant energy scale
for the nonzero field. \edit{Suppression of rare sample could further improve the economy of machine learning methods for spin glasses in a field.}

\edit{In specific} renormalization group (RG) terms, the neural network has
built in a particular representation for the RG flow for only the SG,
FM, and PM states for a spin-glass model at zero field after training,
and by assumption excludes the possibility of an additional fixed point
to cause an additional state for the neural network to consider. While
this would indeed cast the qualitative validity of the neural network's
predictions at nonzero field into question, we believe that this is a
very unlikely scenario for three-dimensional spin glasses where our
results agree qualitatively with multiple numerical studies in the
literature
\cite{caracciolo:90,houdayer:99,young:04,katzgraber:05c,sasaki:07b,joerg:08a,katzgraber:09b,larson:13}. \edit{Adapting our framework to design neural networks for higher-dimensional and long-range spin glass models is an important future step in concretely establishing the range and utility of machine learning in equilibrium spin glass physics.}

In general, NNs once optimized could be used to probe the phase
characteristics of modified spin-glass models where Monte Carlo
statistics may happen to be unavailable or inconclusive. Examples are
systems where no local order parameter exists, such as in lattice gauge
theories found for some error models when determining error thresholds
for topological codes
\cite{dennis:02,arakawa:05,katzgraber:09c,andrist:10,bombin:12,andrist:12}.

Further computational and theoretical research in this disciplinary
interface may be encouraging. Both neural networks and spin glasses are
examples of complex systems with a breadth of applications, where in
fact some neural networks can find a description though long-range spin
glasses \cite{nishimori:01}. The converse study of describing spin
glasses through neural networks may prove rewarding from the point of
view of complexity theory.

\section{Acknowledgments}

We would like to thank Amin Barzegar, Chao Fang and Evert van
Nieuwenburg for comments and useful discussions. H.~M.~B.~acknowledges
financial support from the Marianne '76 and Robert '77 Hamm Endowed
Scholarship from the Department of Physics and Astronomy while a student
at Texas A\&M University. The research of H.~M.~B.~is based upon work
(partially) supported by the Office of the Director of National
Intelligence (ODNI), Intelligence Advanced Research Projects Activity
(IARPA), via the U.S. Army Research Office contract W911NF-17-C-0050.
H.~G.~K.~acknowledge support from the National Science Foundation (Grant
No.~DMR-1151387). The work of H.~G.~K.~is supported in part by the
Office of the Director of National Intelligence (ODNI), Intelligence
Advanced Research Projects Activity (IARPA), via MIT Lincoln Laboratory
Air Force Contract No.~FA8721-05-C-0002. The views and conclusions
contained herein are those of the authors and should not be interpreted
as necessarily representing the official policies or endorsements,
either expressed or implied, of ODNI, IARPA, or the U.S.~Government. The
U.S.~Government is authorized to reproduce and distribute reprints for
Governmental purpose notwithstanding any copyright annotation thereon.
We thank Texas A\&M University for access to their Ada and Terra
clusters.

\bibliography{refs.bib}

\end{document}